\documentclass[12pt]{article}

\usepackage{amssymb}
\usepackage{amsmath}

\usepackage{graphicx}
\usepackage{rotating}

\DeclareGraphicsExtensions{.png}
\DeclareGraphicsExtensions{.jpg}
\DeclareGraphicsExtensions{.tif}
\DeclareGraphicsExtensions{.JPG}
\usepackage{float}

\usepackage{csquotes}

\usepackage{array, booktabs, caption}
\usepackage{makecell}

\usepackage{natbib}

\usepackage{authblk}

\date{\vspace{-5ex}}

\usepackage[T1]{fontenc}
\usepackage{titling}
\usepackage{tocloft}
\cftpagenumbersoff{figure}
\cftpagenumbersoff{table}

\usepackage{hyperref}

\begin{document}

\title{Interactive Estimation of the Fractal Properties of Carbonate Rocks  }
\author{Adewale Amosu$^\star$ $^\dagger$, Hamdi Mahmood$^\dagger$, Paul Ofoche$^\ddag$, Mohamed Imsalem$^\dagger$}

\affil{$^\dagger$Department of Geology and Geophysics,
\\ Texas A$\&$M University, College Station, 
\\Texas, 77843, USA
\\ $^\star$adewale@tamu.edu}

\affil{$^\ddag$Department of Petroleum Engineering,
\\ Texas A$\&$M University, College Station, 
\\Texas, 77843, USA}
\maketitle

\section*{Abstract}
Scale invariance of intrinsic patterns is an important concept in geology that can be observed in numerous geological objects and phenomena. These geological objects and phenomena are described as containing statistically selfsimilar patterns often modeled with fractal geometry.  Fractal geometry has been used extensively to characterize pore space and fracture distribution of both carbonate and clastic rocks as well as the transport properties of porous media and fluid flow in reservoirs. The fractal properties are usually estimated from thin-section photomicrograph images or scanning electron microscope images. For complex rock such as carbonate rocks, automatic feature detection methods are often inaccurate. In addition, the rocks may be have been subjected to facies selective diagenesis which preferentially affect some of the rock fabric, thus increasing the difficulty in automatic detection of certain features.  We present an interactive program, GeoBoxCount, for analyzing thin-section images and calculating the fractal dimension interactively. The program relies on the geologists insight in interpreting the features of interest; this significantly improves the accuracy of feature selection. The program provides two options for calculating the fractal dimension: the Hausdorff and the Minkowsi-Bouligand box-counting methods.

\section{Introduction}
The concept of fractals was introduced by Benoit Mandelbrot (1983) and can be observed extensively in many areas of geology and geophysics. Examples include the perimeter of coastlines (Mandelbrot, 1983), sinuosity of stream patterns, velocity modeling in refraction seismology (Crossley and Jensen, 1989), oil and gas field distributions (Hein, 1999), and in the frequency-intensity distribution of earthquakes (Turcotte, 1992). Fractal geometry has also been used extensively to characterize pore space and fracture distribution of both carbonate and clastic rocks as well as the transport properties of porous media and fluid flow in reservoirs (Pape et al. 1987, Pape et al. 1999, Xie et al. 2010). 
Box-counting methods (Moisy, 2006) are commonly applied to thin-section photomicrographs or scanning electrom microscope (SEM) images in order to estimate the fractal dimension. This procedure involves recognizing every instance of a certain feature everywhere it occurs in the image, then super-imposing boxes of varying size and counting how many boxes cover the features of interest. When feature interpretation is done automatically the recognition accuracy is often low. The procedure is affected by other problems such as: finite size effects (Gonzato et al, 1998), edge effects (Agterberg et al., 1996) and memory limitations (Hou et al. 1990). This has led to the proposal of different box-counting methods.

We present a new program, GeoBoxCount, for interpreting thin-section photomicrograph and SEM images and interactively selecting features of interest based on a geological insight. Due to complexity of certain rocks, such as carbonate rocks that are often affected by facies selective diagenesis, automatic detection of features, such as pore spaces, fractures, or specific fossil types, is often inaccurate. Interactive interpretation of the images ensures all features of interest are captured. After the features have been selected, the program provides two options for calculating the fractal dimension: the Hausdorff and the Minkowski-Bouligand or Kolmogorov box-counting methods. 

\section {GeoBoxCount}
GeoBoxCount is written in Matlab 2017b and includes a graphical user interface (GUI) for running the program (Figure \ref{fig:1} ). The program can be downloaded from \url{https://zenodo.org/record/1174524}. It can be run using the following simple functionalities:

\begin{figure}[!h]
	\centering	\includegraphics[width=15cm,height=10cm]{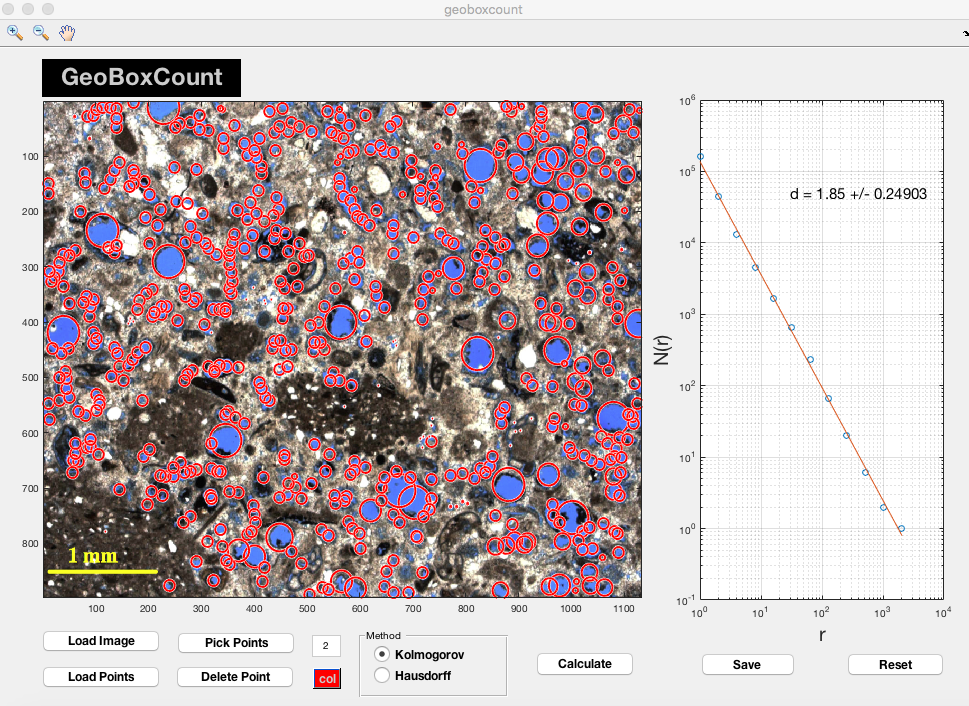}
	\caption{Graphical User Interface for GeoBoxCount}
	\label{fig:1}
\end{figure}

\begin{enumerate}
\item Load Image: The Load Image button is used to input the image to be analyzed. The program accepts most common image formats. The main axis also has zoom functionality.

\item Picking Points: The Pick Points button is used to interpret the image. By default is uses circles. The number panel and the color panel can be used to set the radius and the color of the circle.

\item  Load Points: The Load Points button can be used to input points from an ascii file saved from a previous pick. 

\item Method: The Method Panel is used to select the preferred method for calculating the fractal dimension.

\item Calculate: The Calculate button is used implement the calculation and display the results. 

\item General: The Save button is used to save the images and the picked coordinates. The Reset button, clears the memory and restarts the program.
\end{enumerate}

\section{Application}
\begin{figure}[!h]
	\centering	\includegraphics[width=10cm,height=9cm]{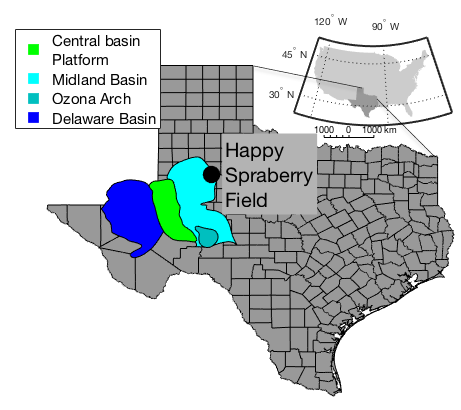}
	\caption{Location of the Happy Spraberry Field, Permian Basin, TX}
	\label{fig:2}
\end{figure}

To demonstrate the usage of the program we apply it to data from the Happy Spraberry Field, Permian basin, TX, USA (Figure \ref{fig:2}). We also demonstrate how the fractal dimension estimated from moldic pores can be used to calculate permeability. 
The Happy Spraberry Field Texas is located in Garza County on the northern part of the Midland Basin (Figure \ref{fig:2}). It produces oil from heterogeneous shallow shelf carbonates of the Permian-aged Lower Clear Fork Formation. Core samples obtained from the field indicate the reservoir facies contain oolitic skeletal grainstones/packstones and skeletal rudstones. The reservoir facies have cemented and dissolution enhanced pore types caused by facies selective diagenesis. Moldic pores are the most abundant across the field and dominate the oolitic skeletal grainstone packstone facies. We make use of thin section photomicrographs of the reservoir facies from a well in the Happy Spraberry Field. We use GeoBoxCount to interactively model the pore paces as tubular cylinders in a pigeon hole fractal model and apply the box-counting method to extract the porosity and the Minkowsi-Bouligand fractal dimension. For a pigeonhole fractal model Pape et al. (1987) and Pape et al. (1999) derived equations that relate tortuosity and porosity with the fractal dimension. Equation (\ref{eq1}) shows the modified Kozeny Carman equation. Equations (\ref{eq2}) and (\ref{eq3}) relate tortuosity and porosity to the fractal dimension; the equations are only valid for fractal dimensions with values between 2 and 3.

\begin{equation}
k=\frac{\phi}{8T}r_{eff}^2
\label{eq1}
\end{equation}
\begin{equation}
T=1.34 \frac{r_{grain}}{r_{eff}}^{0.67(D-­2)}
\label{eq2}
\end{equation}
\begin{equation}
\phi = 0.5 \frac{r_{grain}}{r_{eff}}^{0.39(D-­3)}
\label{eq3}
\end{equation}

In the equations (\ref{eq1} - \ref{eq6}), T is tortuosity, $r_{grain}$ is average grain size, $r_{eff}$ is the effective pore radius, D is the fractal dimension, k is permeability and $\phi$ is porosity.

\begin{figure}[!h]
	\centering	\includegraphics[width=10cm,height=8cm]{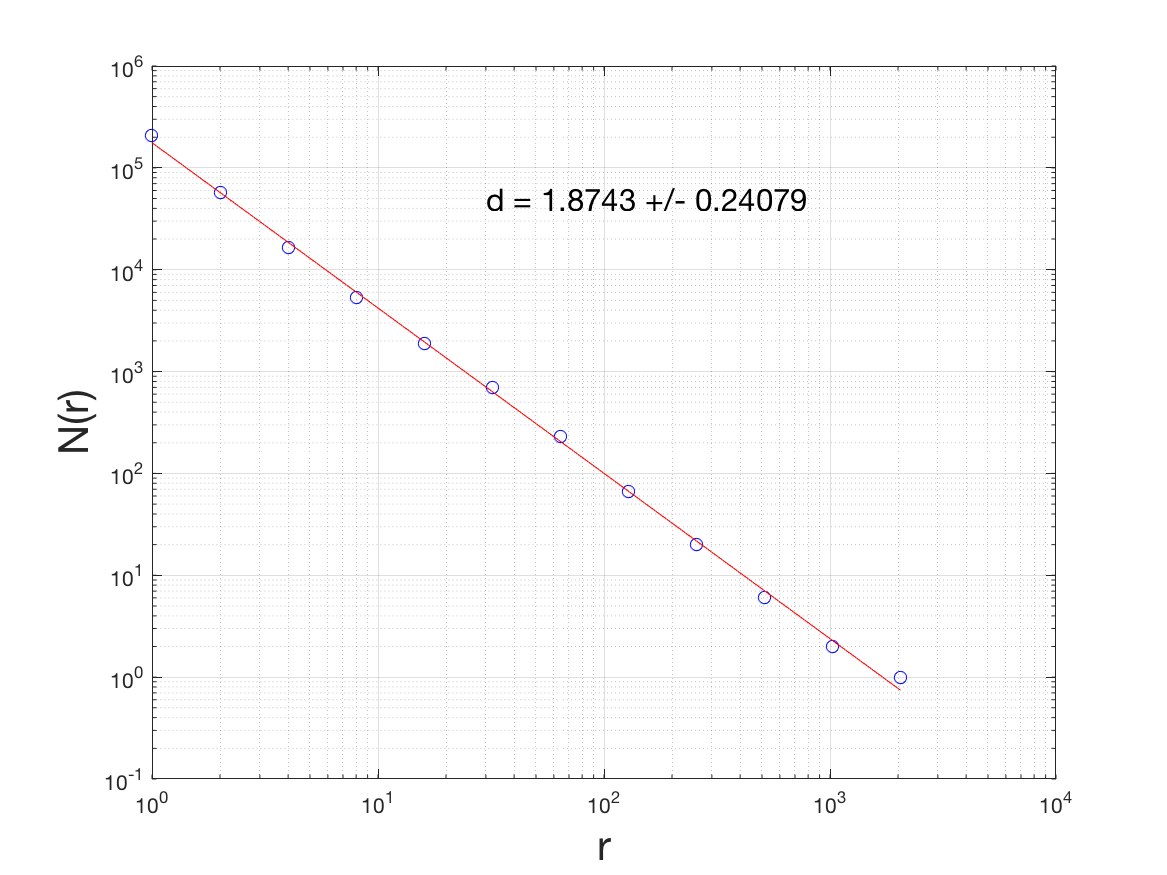}
	\caption{Calculating the fractal dimension}
	\label{fig:4}
\end{figure}

Figure \ref{fig:3} shows the pigeonhole model used to approximate the moldic pores. Figure \ref{fig:4} depicts how the fractal dimension is estimated. Plausible values of the fractal dimension range from 1.63 to 2.11, hence equation (\ref{eq2}) is applicable for values greater than 2. We choose the value of 2.11 and substitute it in the equations  (\ref{eq2}) and (\ref{eq3}) to obtain:
\begin{figure}[!h]
	\centering	\includegraphics[width=12cm,height=9cm]{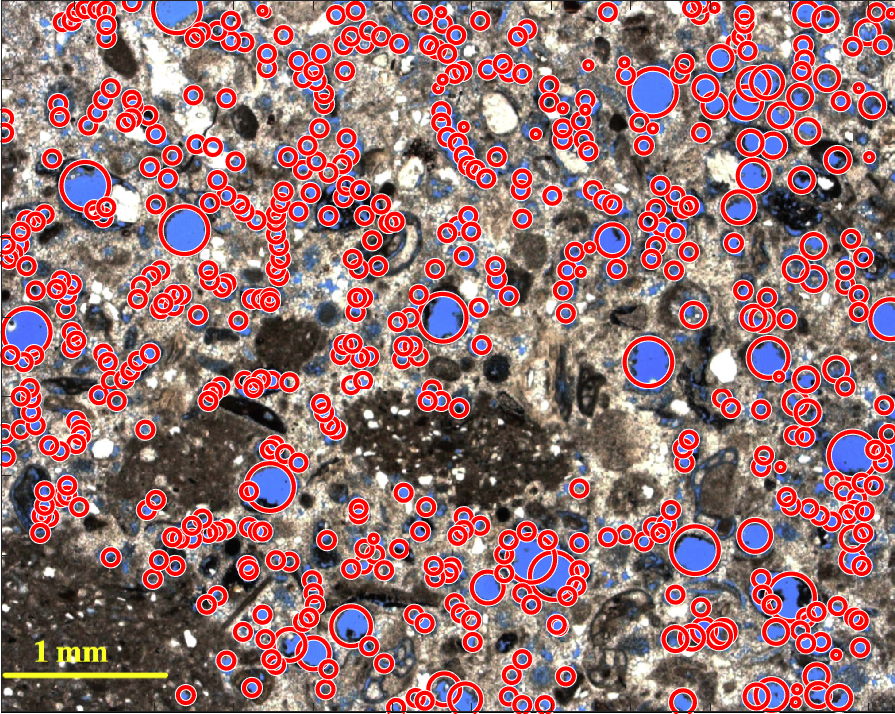}
	\caption{Application of the pigeon-hole fractal model to moldic pores in a thin-section image.}
	\label{fig:3}
\end{figure}
\begin{figure}[!h]
	\centering	\includegraphics[width=10cm,height=8cm]{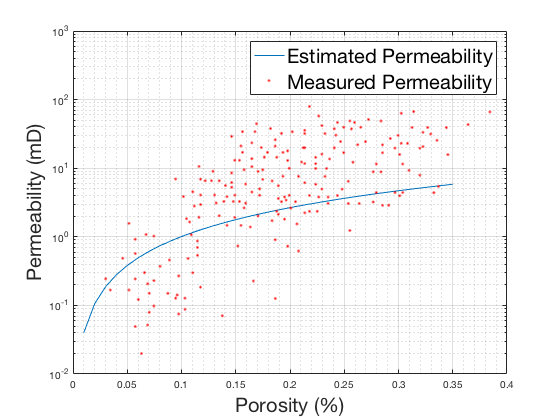}
	\caption{Comparing estimated and measured permeability}
	\label{fig:5}
\end{figure}

\begin{equation}
T=1.34 \frac{r_{grain}}{r_{eff}}^{0.07}
\label{eq4}
\end{equation}
\begin{equation}
\phi = 0.5 \frac{r_{grain}}{r_{eff}}^{-0.35}
\label{eq5}
\end{equation}

Combining equations (\ref{eq4}) and (\ref{eq5}) with equation (1), we obtain equation (\ref{eq6}):
\begin{equation}
k = 4.3 *10^{11} r_{grain}^2 \phi^{7/5}
\label{eq6}
\end{equation}

Using an average value of grain size radius = 250000nm and porosity ranging from 0 to 35 percent, we compare the estimated permeabilities to lab measured core permeabilities from the field. Figure \ref{fig:5} shows the estimated permeability-porosity relationship has a good match with the lab measured permeability-porosity relationship especially for porosity values less than 20 percent.

\section{Conclusion}
We develop a new program for interactively selecting features of interest in thin-section photomicrograph and SEM images and estimate the fractal dimension. We demonstrate the usage of the program using data from the Happy Spraberry Field, Texas. We also demonstrate how to use the fractal dimension in estimating the tortuosity and permeability of rock samples from fractal properties of moldic pores. The pigeonhole fractal is used to successfully characterize the moldic pores in the reservoir facies of carbonate rocks and extract the fractal dimension. We then apply the Kozeny-Carman equation and equations relating the tortuosity and the porosity to the fractal dimension to establish an empirical relationship between permeability and porosity. For more geoscience related codes and works from the authors, see (Attanayake et al., 2010; Amosu et al., 2011, 2012, 2013, 2016; Amosu and Smalley, 2014; Amosu, 2013, 2014; Amosu and Sun, 2017a-e, 2018).

\clearpage
\section*{References}
\begin{enumerate}
\item Mandelbrot, B.B., 1983. The fractal geometry of nature (Vol. 173). New York: WH freeman.

\item Crossley, D. J., and Jensen, O. G., 1989, Fractal velocity models in refraction seismology: Pageoph, 131, 61-76.

\item Hein, F.J., 1999. Mixed ("multi") fractal analysis of granite wash fields/pools and structural lineaments, Peace River Arch area, northwestern Alberta, Canada; a potential approach for use in hydrocarbon exploration. Bulletin of Canadian Petroleum Geology 47, 556-572.

\item Turcotte, D.L., 1992. Fractals and Chaos in Geology and Geophysics. Cambridge University Press, Cambridge 221pp.

\item Pape H., Riepe L., Schopper J., 1987. Theory of self‐similar network structures in sedimentary and igneous rocks and their investigation with microscopical and physical methods. Journal of Microscopy 148 (2): 121-147.

\item Pape, H., Clauser, C. and Iffland, J., 1999. Permeability prediction based on fractal pore-space geometry. Geophysics, 64(5), pp.1447-1460.

\item Xie, S., Cheng, Q., Ling, Q., Li, B., Bao, Z. and Fan, P., 2010. Fractal and multifractal analysis of carbonate pore-scale digital images of petroleum reservoirs. Marine and Petroleum Geology, 27(2), pp.476-485.

\item Moisy, F., 1D, 2D and 3D Box-counting, 2006.

\item Gonzato, G., Mulargia, F., Marzocchi, W., 1998. Practical application of fractal analysis: problems and solutions. Geophysical Journal International 132 (2), 275-282 . doi:10.1046/j.1365-246x.1998.00461.x.

\item Agterberg, F.P., Cheng, Q., Brown, A., Good, D., 1996. Multifractal modeling of fractures in the Lac du Bonnet Batholith, Manitoba. Computers $\&$ Geosciences 22 (5), p 497-507.

\item Hou, X.-J., Gilmore, R., Mindlin, G.B., Solari, H.G., 1990. An efficient algorithm for fast O(N ln(N)) box counting. Physics Letters A 151, 43-46 . doi:10.1016/ 0375-9601(90)90844-E.

\item Amosu A., Y. Sun, WheelerLab: An interactive program for sequence stratigraphic analysis of seismic sections, outcrops and well sections and the generation of chronostratigraphic sections and dynamic chronostratigraphic sections {\it SoftwareX} {\bf 6} (2017), pp 19-24.

\item Amosu A., Y. Sun, FischerLab: An Interactive Program for Generating Dynamic Fischer Plots From Wireline Logs and Stratigraphic Data, {\it AAPG Annual Convention } (2017).

\item Amosu A., Y. Sun, WheelerLab: An Interactive Program for Sequence Stratigraphic Analysis of Seismic Sections and the Generation of Dynamic Chronostratigraphic Sections, {\it AAPG Annual Convention } (2017).

\item Amosu A., Y. Sun, FischerLab: An interactive program for generating Fischer plots and stepwise Fischer plots from wireline logs and stratigraphic data: SoftwareX (Under Review).

\item Amosu A., Y. Sun, Visualization of Angular Unconformities and Tectonic Angular Discordance Measurement Constraints by Structural Geometrical Flattening: Case Studies in the Permian (California), Grand Canyon (Arizona), Chad basin (Nigeria), Algarve Basin (Iberia) and the Aegean Sea Basin (Turkey), {\it Gulf Coast Association of Geological Societies Transactions} (2017).

\item Amosu A., Y. Sun,Sequence Stratigraphy, Chronostratigraphy and Spatio-Temporal Stratigraphic Thickness Variation of the Agbada Formation, Robertkiri and Delta Fields, Niger Delta, Nigeria, {\it Gulf Coast Association of Geological Societies Transactions} (2017).

\item Amosu, A. and Mahmood, H., 2018. PyLogFinder: A Python Program for Graphical Geophysical Log Selection. Research Ideas and Outcomes, 4, p.e23676.

\item Amosu, A., 2014. Elastic Deformation of the Earth's Crust from Surface Loading Phenomena (Doctoral dissertation, University of Memphis).

\item Amosu, A. and Sun, Y., 2018. MinInversion: A Program for Petrophysical Composition Analysis of Geophysical Well Log Data. Geosciences, 8(2), p.65.

\item Attanayake, J., Ghosh, A. and Amosu, A., 2010, December. Introductory Earth science education by near real time animated visualization of seismic wave propagation across Transportable Array of USArray. In AGU Fall Meeting Abstracts.

\item Amosu, A., Smalley, R., Wilson, T.J., Bevis, M.G., Dalziel, I.W., Kendrick, E.C., Konfal, S., Magee, W.R. and Stutz, J.E., 2011, December. Automatic Processing of Antarctic Polenet GPS Data. In AGU Fall Meeting Abstracts.

\item Amosu, A., Smalley, R. and Puchakayala, J., 2013. Modeling Earths Crustal Deformation In the Lower Mississippi River Basin. Seismological Research Letters, 84(2), p.376.

\item Amosu, A., 2013. Parallel Computation Algorithm (PCA) for Large Geophysical Processes. American Association of Petroleum Geologists, AAPG/SEG Expo, Article, 90182.

\item Amosu, A. and Smalley, R., 2014. Crustal Deformation from Surface Loading In the Great Salt Lake Region. Seismological Research Letters, 85(2), p.447.

\item Amosu, A., Sun, Y. and Agustianto, D., 2016, January. Coherency-based inversion spectral decomposition of seismic data. In 2016 SEG International Exposition and Annual Meeting. Society of Exploration Geophysicists.

\item Amosu, A., Smalley, R. and Puchakayala, J., 2012, December. Modeling Earth deformation from the 2011 inundation in the Mississippi river basin using hydrologic and geodetic data. In AGU Fall Meeting Abstracts.

\end{enumerate}
\end{document}